\documentclass[aps,pre,twocolumn,showpacs,superscriptaddress,groupedaddress]{revtex4-2}
\usepackage{amsmath}
\usepackage{amsfonts}
\usepackage{amssymb}
\usepackage{graphicx}
\usepackage{graphics}
\usepackage{dcolumn}
\usepackage{sidecap}
\usepackage{parskip}
\usepackage{xcolor}
\usepackage{hyperref}
\usepackage{hhline}
\usepackage{mathtools}
\usepackage{multirow}
\usepackage{verbatim}
\usepackage{rotating}
\usepackage{setspace}
\usepackage{epsfig}
\usepackage{epstopdf}
\usepackage{natbib}
\usepackage{subfigure}
\usepackage{booktabs}
\usepackage[normalem]{ulem}
\usepackage{bm}
\usepackage{comment}
\usepackage{verbatim}
\usepackage[normalem]{ulem}
\usepackage{comment}
\usepackage{cleveref}
\usepackage[normalem]{ulem} 
\usepackage{lineno}

\makeatletter
\def\@eqnnum{{\normalsize \normalcolor (\theequation)}}
 \makeatother
\makeatletter
\def\section{\@startsection {section}{1}{\z@}%
  {-3.5ex \@plus -1ex \@minus -.2ex}%
  {2.3ex \@plus.2ex}%
  {\normalfont\bfseries\raggedright}} % numbered section: bold left aligned no uppercase

\def\section*#1{%
  \par\addvspace{2.3ex \@plus.2ex}%
  \noindent
  {\normalfont\bfseries #1}%
  \par\nobreak\addvspace{1.5ex \@plus .2ex}%
}
\makeatother
\graphicspath{{./}{ER/main/}}
\usepackage{tikz} % Needed for drawing lines
\begin{document}
\title{Second-order Kuramoto model with adaptive simplicial complex}
\author{Priyanka Rajwani and Sarika Jalan}
\email{Corresponding Author: sarika@iiti.ac.in} % The corresponding author email should go here.
\affiliation{Complex Systems Lab, Department of Physics, Indian Institute of Technology Indore, Khandwa Road, Simrol, Indore-453552, India}
%\linenumbers
\begin{abstract}

We investigate the emergence of synchronization in the second-order Kuramoto model with adaptive simplicial interactions on a globally connected network. 
%We address this by analyzing synchronization in the second-order Kuramoto model with adaptive simplicial interactions on a globally connected network. 
This inertial Kuramoto framework describes systems, where oscillator frequencies evolve over time. Unlike most previous work that ignores inertia, we examine how inertia combined with adaptive higher-order coupling alters synchronization transitions. Using self-consistency analysis, we derive the steady-state behavior and show that adaptation qualitatively reshapes the synchronization landscape. We find that the backward transition from synchronization to incoherence remains controlled by the adaptive feedback parameter, but the forward discontinuous jump to synchronization vanishes in the thermodynamic limit. In contrast, finite-size systems still display an abrupt transition to synchronization, with its onset precisely set by the adaptation control parameter. These results show how adaptive feedback and system size together govern the onset and robustness of synchronization in inertial oscillator networks with higher-order interactions.

\end{abstract}   

\maketitle
\section*{{Introduction}}

Synchronization is a fundamental phenomenon observed in various real-world systems, including neural networks, circadian clocks, chemical oscillators, and power grids \cite{pikovsky2003}. 
To understand the origins of synchronization in spatially extended systems, researchers often use the Kuramoto model \cite{kuramoto1975, strogatz2000}. 
This model helps elucidate the mechanisms and dynamical origins of collective behaviors. 
Unlike the classical Kuramoto model, { which describes overdamped phase dynamics}, the Kuramoto model with inertia \cite{tanaka1997, olmi2014, kati2025} {incorporates an inertial term proportional to the second time derivative of the phase and is commonly referred to as the second-order Kuramoto model. The presence of inertia leads to} a first-order transition to (de)synchronization, accompanied by hysteresis \cite{gao2018}. { Such phase dynamics with inertia arise in systems where instantaneous frequencies evolve dynamically due to interactions; for example, in southeast asian firefly {\textit{Pteroptyx malaccae}}, individuals can adjust their intrinsic flashing frequency in response to external signals \cite{tanaka1997_D, ermentrout1991}.}
This framework is also relevant to power grid dynamics, where stable operation requires that all components remain synchronized at a common frequency \cite{rohden2012, schafer2018, schafer2018b, yang2017}.  
Here, the second-order Kuramoto model naturally arises from the consideration of power conservation at each node within power grid networks, assuming negligible ohmic losses and a high-voltage transmission grid  \cite{filatrella2008, manik2014}.

\begin{figure*}
%\vspace{-2cm}
\includegraphics[width=\textwidth]{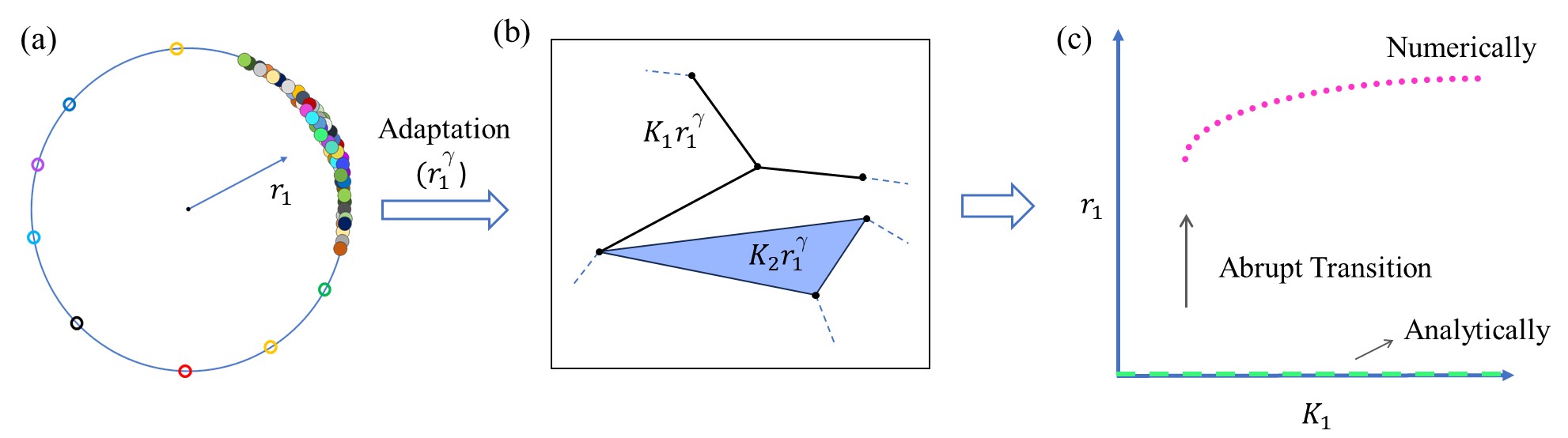}
%\vspace{-4cm}
\caption{Schematic diagram illustrating a dynamical snapshot of the second-order Kuramoto model with higher-order interactions and adaptive coupling. (a) Solid and open circles denote locked and drifting oscillators, respectively. (b) $K_1$ and $K_2$, the coupling strengths for $1$- and $2$-simplex interactions { for globally coupled network}, respectively. {{Where adaptation is governed by the global order parameter $r_1$ (Eq.~\eqref{order_par})}, with the exponent $\gamma$ controlling the strength of the adaptive feedback}. (c) Dynamics of coupled second-order Kuramoto oscillators with adaptive coupling and higher-order interactions (Eq.~\eqref{GC_Model_Eq}). { The schematic depicts that, in finite-size numerical simulations, the forward continuation in $K_1$ exhibits an abrupt jump towards the synchronized state. In contrast, the analytical results, corresponding to the continuum limit ($N \to \infty$), demonstrate the stability of the incoherent state as $K_1$ increases.}}
\label{0}
\end{figure*}
Previous studies on the second-order Kuramoto model have typically considered a constant coupling strength.  

{An adaptive coupling arrangement, in which the interaction strength is modulated by a feedback from the collective synchronization measure.}   Filatrella {\it{et. al.}}  introduced such an adaptation scheme for pairwise interactions in the classical Kuramoto oscillators (without inertia) to study synchronization of Josephson junction arrays coupled through a resonator  \cite{filatrella2007}.  {They demonstrated that increasing the adaptive feedback strength induces a change in the nature of the synchronization transition, from continuous to first order.} Subsequently, a similar adaptive coupling framework for the {{classical}} Kuramoto model was explored analytically in several studies \cite{zhang2015, zou2020, jin2023}. { Moreover, adaptive Kuramoto-type models have also been used to describe collective phenomena such as clapping audiences \cite{taylor2010}.} In the real-world system, adaptation mechanisms can differ based on the network architecture and learning algorithms.  In power grid networks, adaptation may occur through variations in both the network topology and dynamical behaviors of the grid \cite{sawicki2023}. The structural adaptability of networks has been studied in both neuronal and power grid systems \cite{berner2021, kachhvah2022, fialkowski2023}.

Furthermore, in many real-world systems, interactions often extend beyond pairs and frequently occur in groups.  These higher-order interactions have been shown to play crucial roles in the dynamical evolution of the underlying system, influencing collective behavior in ways that cannot be captured by pairwise interactions alone \cite{tanaka2011, battiston2020, boccaletti2023, anwar2024, dong2025}. When higher-order interactions, often encoded through simplicial complexes, are incorporated into the Kuramoto model, they show a significant impact on the coupled dynamical behavior by shifting transition points of the system, more importantly, introducing new collective states \cite{skardal2019, skardal2020, jalan2022}. Additionally, in the Kuramoto model without inertia, the introduction of adaptive higher-order coupling, {modulated by a measure of collective synchronization}, can lead to an abrupt to continuous jump to a synchronized state or to tiered synchronization \cite{rajwani2023, dutta2024}. For the Kuramoto model with inertia, only a few recent studies have examined the impact of higher-order interactions \cite{jaros2023, sabhahit2024, rajwani2025, lourenco2025}. { The role of adaptive coupling within this framework remains unexplored.}
 
This study proposes an analytically tractable {theoretical framework based on} the second-order Kuramoto model with higher-order interactions and an adaptive coupling scheme, { in which adaptation is introduced through feedback from the measure of global synchronization.} By employing a self-consistency method in the continuum limit ($N \to \infty$), we examine the behavior of the global order parameter (a measure of phase synchronization) and demonstrate that starting from an incoherent state, an increase in coupling strength allows adaptive interactions to prevent a sudden transition to synchronization. {This is in contrast with the constant-coupling case, where an increase in pairwise coupling strength leads to a first-order transition to synchronization in the thermodynamic limit \cite{sabhahit2024}.} 
However, numerical simulations for the adaptive coupling scheme considered here reveal that an abrupt transition can occur due to the finite size of the system.  We investigate how the transition points corresponding to abrupt jumps to and from the synchronized states depend on several factors, including system size, inertia, the adaptation exponent, higher-order coupling strength, and additive noise.  Our results indicate that a change in the adaptation exponent influences both the forward and backward transition points associated with the abrupt jump from an incoherent to a synchronized state and vice versa.  Also, we analyze how adaptive coupling affects the behavior of multistable states arising due to the presence of inertia.

\section*{Results}
\paragraph{Model} { The second-order Kuramoto model considered in this work naturally emerges from power conservation at each node of a power grid network under standard simplifying assumptions \cite {filatrella2008}. We further extend this framework by incorporating a $2-$simplex (triadic) interaction term, formulated in analogy with the classical Kuramoto model. A triadic interaction term of this form can be derived through phase reduction of the mean-field complex Ginzburg-Landau equation \cite{leon2019}.} The equation of motion for $N$ globally coupled second-order Kuramoto model that incorporates an adaptive coupling scheme, in which the interaction strength is modulated by feedback from the measure of global synchronization ({\it i.e.}, representing the {extent of phase clustering on the unit circle, as illustrated in Fig.~\ref{0}(a)}), is given by
\begin{figure*}[t!]
\begingroup
\begin{tabular}{cc}
 \includegraphics[width=0.375\textwidth]{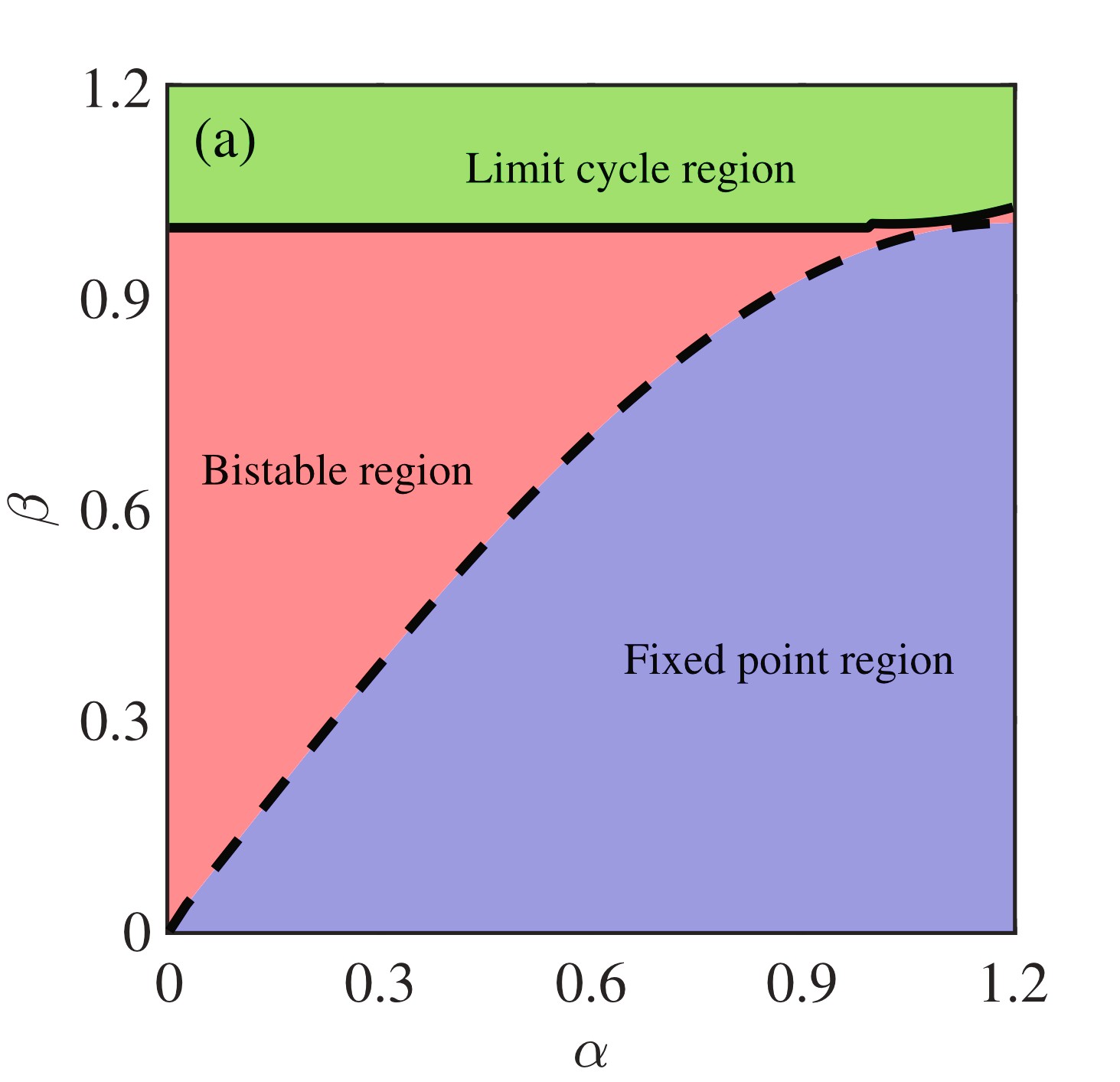}
 \hspace{-0.5cm}
\includegraphics[width=0.37\textwidth]{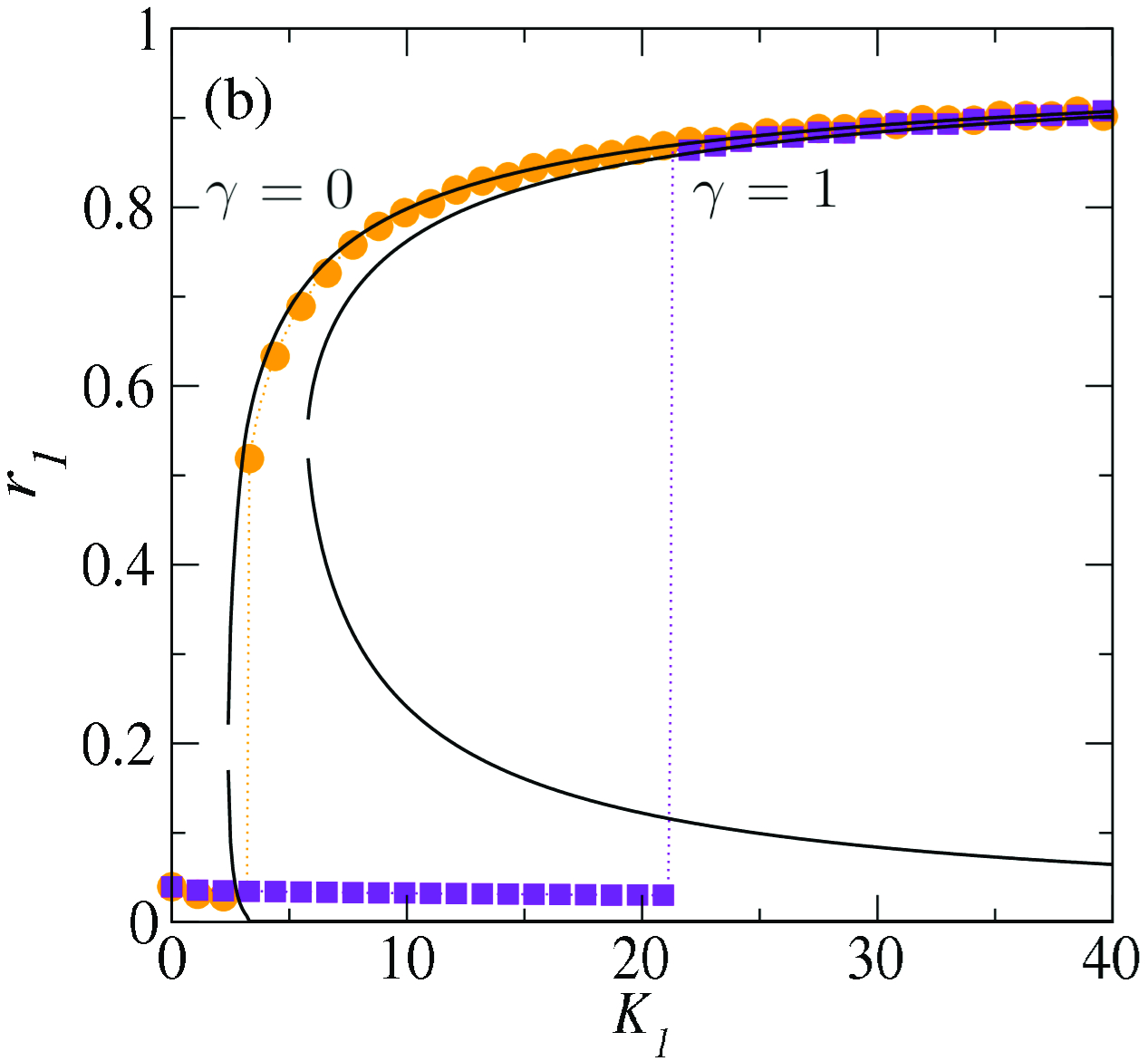}   
\end{tabular}
\endgroup
%\begingroup
\begin{tabular}{c}
 \includegraphics[width=0.72\textwidth]{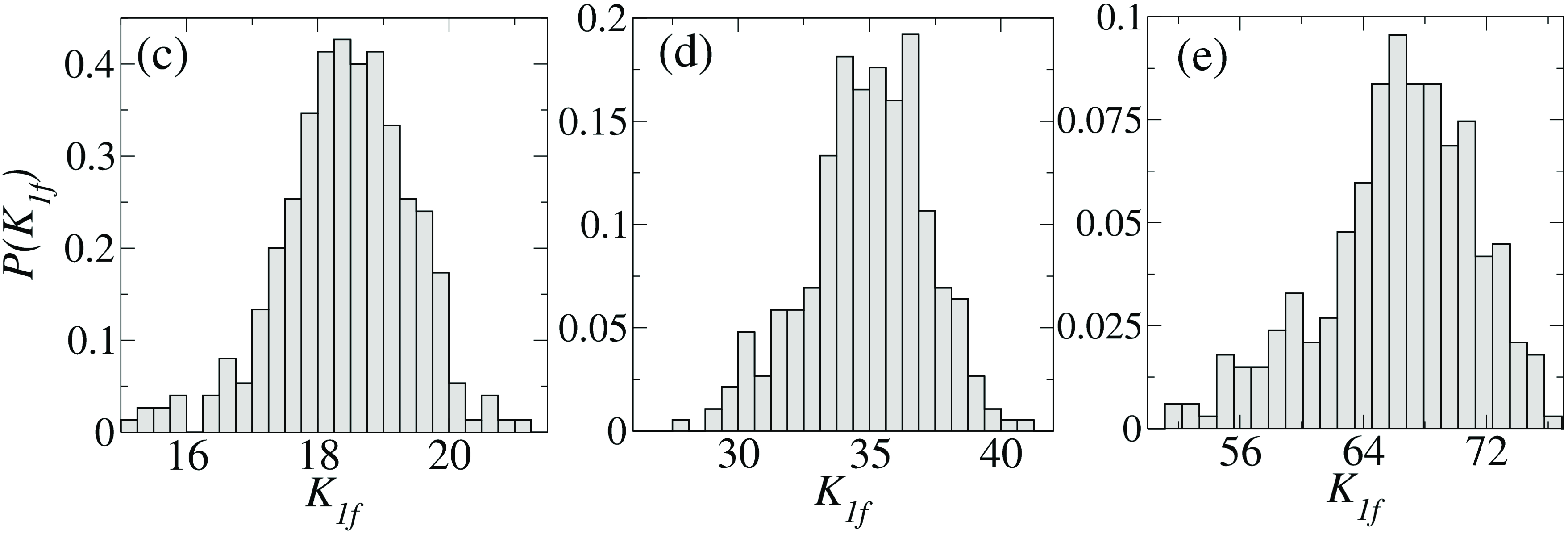}
 %{fin_distribution.eps}
 %\hspace{-0.2cm}
%\includegraphics[width=0.25\textwidth]{dist_al2.eps}  
\end{tabular}
\caption{(Color online) (a) Phase diagram in $\alpha$-$\beta$ plane obtained using Eq.~\eqref{time_MF}, { showing the region of fixed point and limit cycle solutions.} (b) Synchronization profile $r_1$ vs $K_1$ at $K_2=0$ and $m=1$ depicting a jump at $K_{1f}$ due to the finite number of oscillators ($N=500$, and $\gamma=1$). Numerical results for $\gamma=0$ (orange circle), $\gamma=1$ (violet square) are determined considering forward direction simulations using Eq.~\eqref{Mean_field}. Analytical results {for $|\Omega_\mathrm{I}| < \frac{4}{\pi} \sqrt{\frac{q}{m}}$} are shown by solid lines derived using Eq.~\eqref{r1_eq}. {(c-e) Probability distribution of transition point} $K_{1f}$ for (c) $\gamma=1$, (d) $\gamma=1.5$, and (e) $\gamma=2$, calculated for $300$ different realizations {of randomly distributed initial phases}, by fixing $\Delta K_1=0.2$ at $m=1$, $K_2=0$, and $N=500$, depict a shift in mean $K_{1f}$ toward higher values with increasing $\gamma$. }
\label{1}
\end{figure*}
 \begin{equation}\label{GC_Model_Eq}
\begin{split}
   m\ddot{\theta_i}=-\dot{\theta_i}&+{\Omega_i}+\frac{K_1r_{1}^{\gamma}}{N}\sum_{k=1}^{N}\sin(\theta_k-\theta_i)\\&+\frac{K_2r_{1}^{\gamma}}{N^2}\sum_{k=1}^{N}\sum_{l=1}^{N}\sin(2\theta_k-\theta_l-\theta_i),
 \end{split}   
\end{equation}
 where the generalized complex order parameter is defined as:
 \begin{equation}\label{order_par}
     z_p={r_pe^{\iota\psi_p}=\frac{1}{N}\sum_{k=1}^{N}e^{\iota p \theta_k}}.
\end{equation}
 { For $p \in \{1,2\}$, $z_1$ and $z_2$ represent the centroids of the $N$ points $e^{\iota \theta_k}$ and $e^{\iota 2\theta_k}$, respectively, in the complex plane. The magnitude $r_1$ ($0 \le r_1 \le 1$) quantifies the degree of global synchronization in the oscillator system, while the corresponding phase $\psi_1$ denotes the mean phase of the oscillators. In contrast, $r_2$ denotes the order parameter associated with the second harmonic of the phase and becomes prominent in two-cluster configurations where oscillators are separated by a phase difference of $\pi$ \cite{skardal2019}.} The case $r_1 = r_2 = 0$ corresponds to a fully incoherent state, whereas $r_1 = 1$ indicates complete global synchronization, and $r_2>r_1$ represents two-cluster synchronization. $\theta_i$ and $\Omega_i$ denote the phase and intrinsic frequency, respectively, of the $i-$th oscillator { ($i \in \{1,2,\ldots,N\}$), where $\Omega_i$ is drawn from a Lorentzian distribution $g(\Omega)=\frac{\sigma}{\pi[(\Omega-\langle\Omega\rangle)^2+\sigma^2]}$, with mean $\langle \Omega\rangle=0$ and standard deviation $\sigma=1$.} Parameters $K_1$ and $K_2$ denote the coupling strengths for the $1-$ (pairwise) and $2-$ simplex (triadic) interactions { in globally coupled network}, respectively, and $m$ is the inertia.  {
 %Furthermore, the exponent $\gamma$ controls how strongly the coupling adapts through $r_1$ in both the pairwise and triadic interactions.
 In Eq.~\eqref{GC_Model_Eq}, the adaptive mechanism is introduced by allowing the coupling strengths $K_1$ and $K_2$ to be modulated by $r_1^{\gamma}$. Where $\gamma > 0$ is a positive adaptation exponent and determines how strongly the coupling strength associated with pairwise and triadic interactions depends on $r_1$, and thereby regulates the strength of the adaptive feedback.}

Further, using the definition of the order parameter given in Eq.~\eqref{order_par}, Eq.~\eqref{GC_Model_Eq} can be rewritten in a mean-field form. {This is obtained by multiplying Eq.~\eqref{order_par} by $e^{-\iota\theta_i}$, taking the imaginary part, and substituting the resulting expression into Eq.~\eqref{GC_Model_Eq}. In this representation, each oscillator does not interact directly with individual oscillators but rather through the mean-field quantities.} Accordingly, Eq.~\eqref{GC_Model_Eq} takes the mean-field form:

  \begin{equation}\label{Mean_field}
  \begin{split}
   m\ddot{\theta_i}=-\dot{\theta_i}&+{\Omega_i}+{K_1r_{1}^{\gamma+1}}\sin(\psi_1-\theta_i)\\&+{K_2r_{1}^{\gamma+1}r_{2}}\sin(\psi_2-\psi_1-\theta_i).
 \end{split}   
\end{equation}
%{\color{red}This model can be interpreted as a reduced form of the swing equation governing power grid dynamics, which follows from power conservation at each node. Assuming constant voltage amplitudes and sinusoidal coupling between phases \cite{filatrella2008}.}
\paragraph{Analytical results:} 
{In the phase-locked state, a single group of oscillators is formed that is locked to the mean phase $\psi_1$ and rotates uniformly with a common angular velocity $\Omega_0$.} We begin by applying the transformation $\theta_i \rightarrow \theta_i + \Omega_0 t$ in Eq.~\eqref{Mean_field}, 
thereby moving to a rotating frame with frequency $\Omega_0$. {For the synchronized (one-cluster) state analyzed in this work, phases $\psi_1$ and $\psi_2$ are not independent but remain locked in the rotating frame.} Therefore, $\psi_1$ and  $\psi_2$ can be set to zero and Eq.~\eqref{Mean_field} can be written as

\begin{equation}\label{rot_Mean_field}
    m\ddot{\theta_i}=-\dot{\theta_i}+{\Omega_i}-q\sin {\theta_i},
\end{equation}
where $q=r_{1}^{\gamma+1}(K_1+K_2r_2)$. { Here, the index $i$ labels different oscillators only through their intrinsic frequencies $\Omega_i$.} %while all oscillators obey the same functional form of the dynamical equation.} 
By changing the time scale of the system to $\tau = \sqrt{\frac{q}{m}} t$ and dropping the index $i$, Eq.~\eqref{rot_Mean_field} can be written as
\begin{equation}\label{time_MF}
    \ddot {\theta} = - \alpha \dot \theta + \beta - \sin(\theta),
\end{equation}
where $ \alpha = \sqrt{\frac{1}{qm}}$ and $\beta = \frac{\Omega}{q}$. { This reduction reflects the mean-field character of the model, where the collective behavior can be understood by analyzing a single oscillator equation and subsequently integrating over the distribution of intrinsic frequencies.} Furthermore, Eq.~\eqref{time_MF} is similar to the equation describing the Josephson junction \cite{strogatz_book}.  To study the bifurcation behavior, we analyze the phase space plotted in the $\alpha$-$\beta$ plane (Fig.~\ref{1}(a)). The fixed points of the system are found by setting $\ddot{\theta} = \dot{\theta} = 0$, which leads to $\beta = \sin(\theta^*)$. For $\beta < 1$, the system has two fixed points: a saddle and a sink \cite{strogatz_book}. For $\beta > 1$, no fixed point exists, and the system settles to a stable limit cycle. {Following previous studies, phase-space analysis shows that when the system is initialized from an incoherent state, the transition to the synchronized state occurs through a homoclinic bifurcation at $\beta \approx \frac{4}{\pi}\alpha$ depicted as dashed curve in Fig.~\ref{1}(a), as obtained using Melnikov’s method. At this bifurcation point, the limit cycle collides with a saddle equilibrium, forming a homoclinic orbit. In contrast, when the system is initialized from a fully synchronized state, the stable fixed point disappears for $\beta > 1$, leading the system back to incoherent dynamics \cite{gao2018, sabhahit2024}.} In the bistable region depicted in Fig.~\ref{1}(a), a stable fixed point and a stable limit cycle coexist. Moreover, we use the self-consistency method (see ``Methods" for details) to study the dynamics of Eq.~\eqref{GC_Model_Eq} in low-dimensional form as $r_p=r_p^l+r_p^d$, in which the contribution from locked and drifting oscillators is given as
\begin{equation}\label{r1_eq}
%\begin{split}
r_p=  2\left(\int_{0}^{\Omega_{\mathrm{I}}/\Omega_{\mathrm{II}}} \cos(p\theta^*)
+ \int_{\Omega_{\mathrm{I}}/\Omega_{\mathrm{II}}}^{\infty} \langle \cos (p\theta) \rangle\right) g(\Omega) d\Omega,
%\end{split}
\end{equation}
where $ \langle \cos\theta\rangle=-\nu_0^2+\nu_0\sqrt{\nu_0^2-\delta^2}
    $, and $\langle\cos2\theta\rangle=(\nu_o^2\delta^2-\alpha^2\delta^2)\left(\frac{2\nu_0}{\delta^2}\left(\nu_0-\sqrt{\nu_0^2-\delta^2}\right)-1\right)$ with ($\nu_0=\frac{\beta}{\alpha}, \,\, \frac{1}{\delta}e^{\iota\theta^*}=\nu_0+\iota\alpha$).  Additionally, the frequency limit  $|\Omega_\mathrm{I}| < \frac{4}{\pi} \sqrt{\frac{q}{m}}$ and $|\Omega_\mathrm{II}|<q$ correspond to case ($\mathrm{I}$), and case ($\mathrm{II}$), respectively, determined from the phase-space analysis for forward and backward initialization protocols (see ``Methods"). We observe different states based on $r_1$ values, the upper branch solution of Eq.~\eqref{r1_eq} with frequency limits $\Omega_\mathrm{I}$ and $\Omega_\mathrm{II}$ corresponds to weakly and completely synchronized states, respectively, while small $r_1$ values correspond to the lower branch solution. Further, we evaluate the integration in Eq.~\eqref{r1_eq} for ($p=1$) considering oscillators lying within frequency limit $|\Omega_\mathrm{I}| < \frac{4}{\pi} \sqrt{\frac{q}{m}}$ with $K_2=0$. We observe that {{for $\gamma=1$}} the lower branch solution (corresponding to small values of $r_1$ in Fig.~\ref{1}(b) depicted by a solid line) 
asymptotically approaches {to incoherent state as $N \to \infty$, indicating that the incoherent state remains asymptotically stable and no transition from incoherent to the synchronized state occurs in the continuum limit.} In contrast, for $\gamma=0$, the lower branch solution of Eq.~\eqref{r1_eq} converges to the incoherent state, { reflecting the presence of a first-order transition in continuum limit at critical coupling $K_{1f}=2(m+1)$, as derived by Sabhahit {\it{et. al.}} \cite{sabhahit2024}}. However, numerical simulations for a finite number of oscillators $N=500$ and $\gamma=1$, (violet square) in Fig.~\ref{1}(b) depicts that at { transition point} $K_{1f}$, the incoherent state loses its stability, and the system exhibits transitions to the synchronized state. {In the forward simulations, small variations in the estimated value of $K_{1f}$ may result from numerical discretizations; accordingly, a fixed time step ($dt = 0.1$) is considered.}

\begin{figure*}[t!]
\begingroup
  \begin{tabular}{cc}
\includegraphics[width=0.375\textwidth]{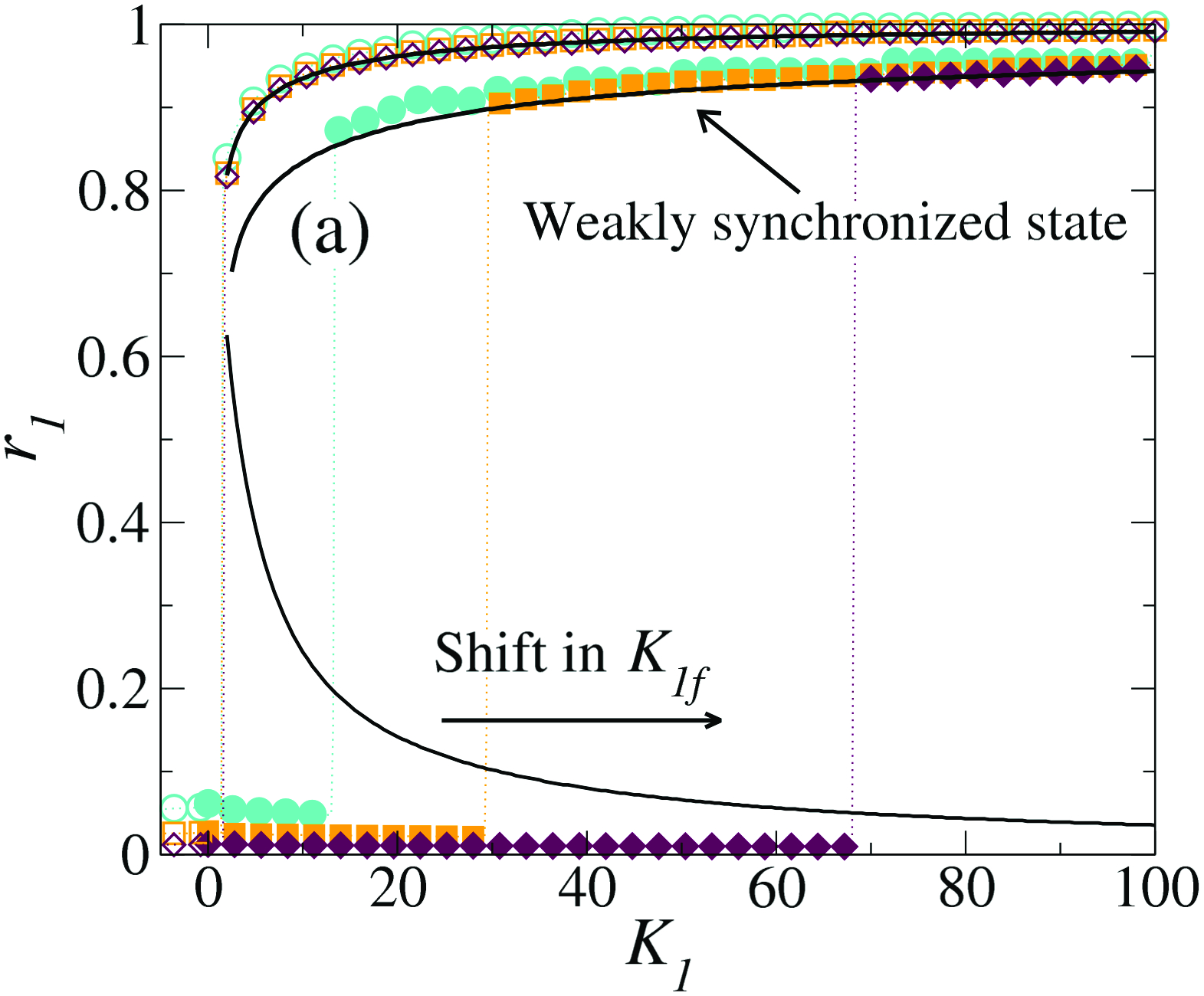} 
\includegraphics[width=0.398\textwidth]{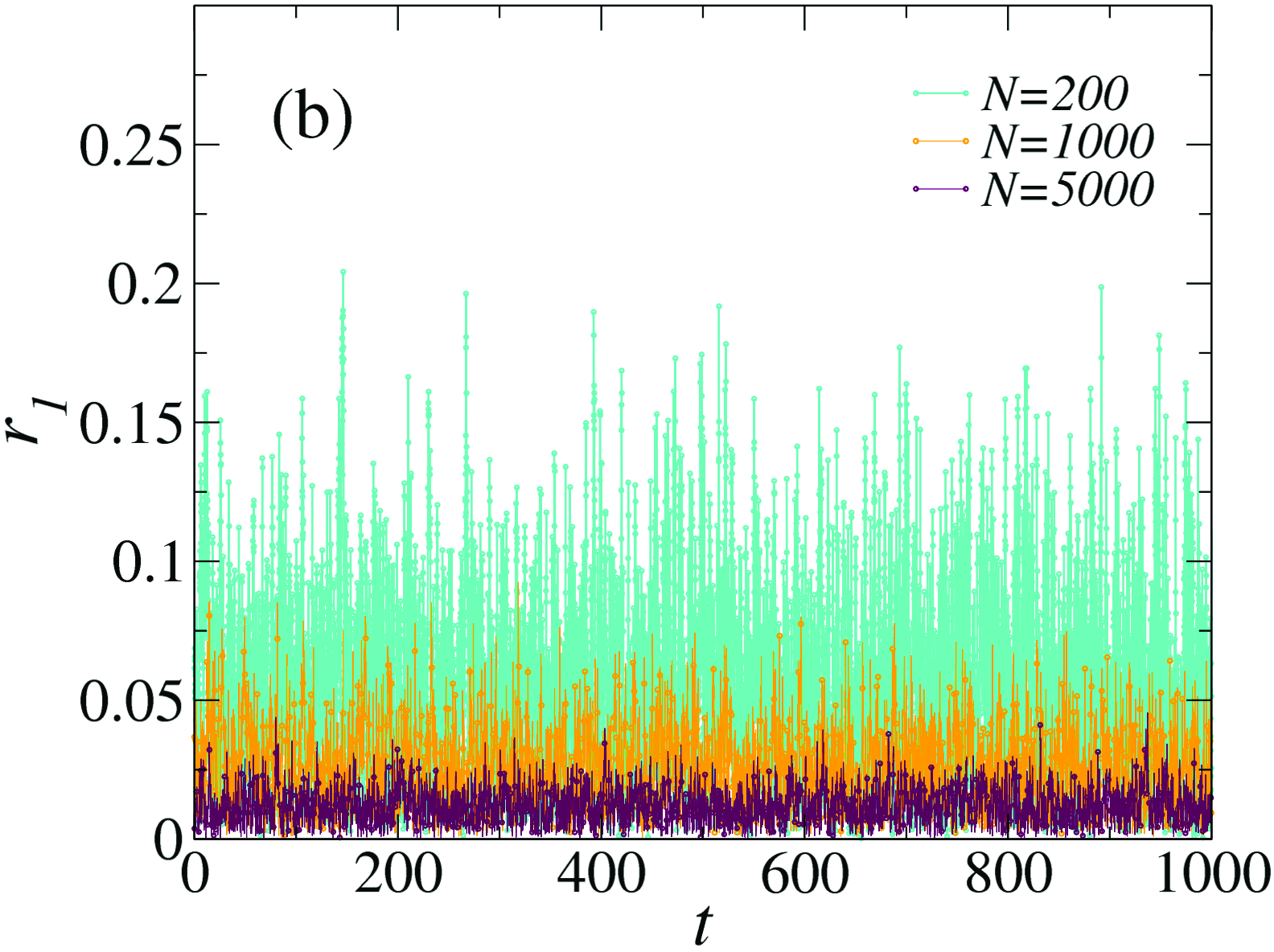}\\
%\hspace{-0.6cm}
\includegraphics[width=0.39\textwidth]{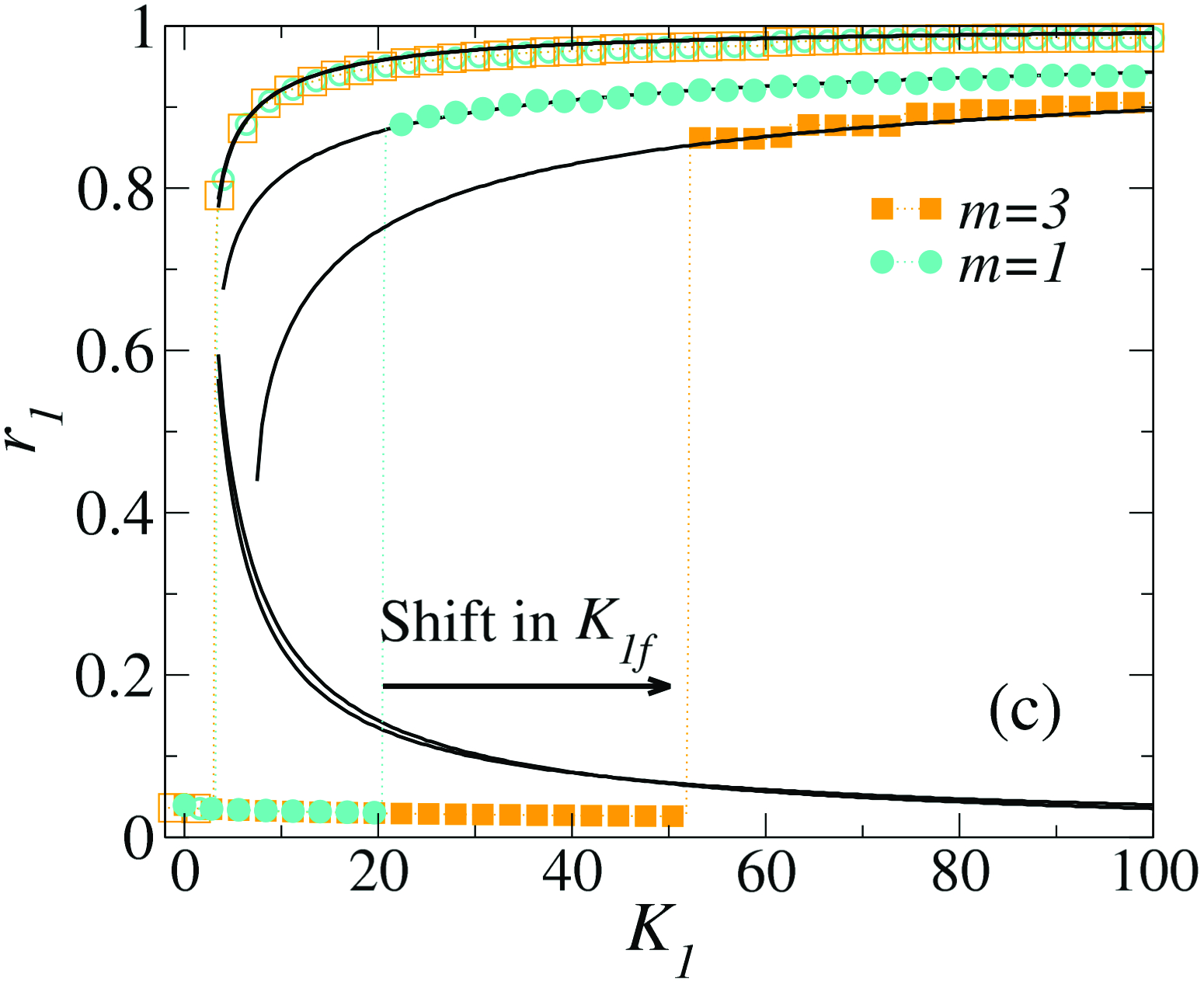} 
\hspace{0.9cm}
\includegraphics[width=0.325\textwidth]{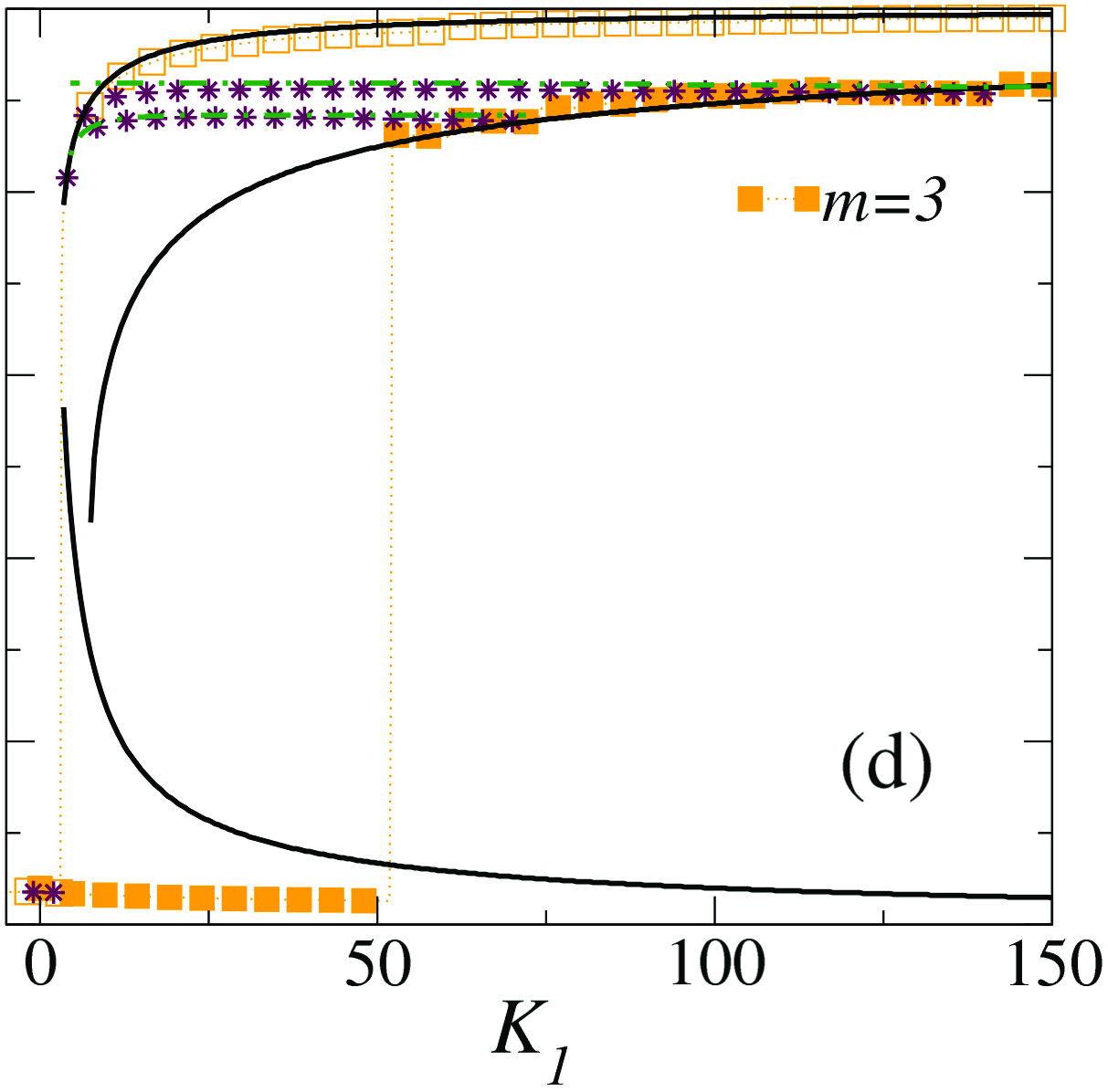} 
\end{tabular}
\endgroup
\caption{(Color online) $r_1$ vs $K_1$ plot for $\gamma=1$. (a) {For fixed  $m=1$ and $K_2=8$, showing the effect of system size} $N=200$ (turquoise circle), $N=1000$ (orange square), $N=5000$ (maroon diamond). { (b) Time evolution of $r_1$ for finite time considering the same parameter values as in panel (a), illustrating the dependence on system size.} (c) For $N=500$ and $K_2=5$, {illustrating the role of inertia} with $m=1$ (turquoise circle), $m=3$ (orange square). (d) Multistability depicted {with (maroon stars)} for $m=3$, $K_2=5$, and $N=500$. The green dot-dashed curves correspond to analytical predictions obtained from Eq.~\eqref{r1_eq} for frequency limit $\Omega = 5.76$ and $8.2$ (from bottom to top). Numerical results are obtained using Eq.~\eqref{Mean_field}, where backward and forward directions are represented by open and filled symbols,
respectively. Analytical results are represented as solid lines determined using Eq.~\eqref{r1_eq}.} 
\label{2}
\end{figure*}
\paragraph{Numerical simulations:}
\label{para:Num_simulation}
Numerical simulations are performed for Eq.~\eqref{Mean_field} using the fourth-order Runge–Kutta (RK-4) method with a time step $dt = 0.1$, by varying $K_1$. In the forward direction, by taking the initial condition as the uniform distribution of $\theta \in (-\pi, \pi)$, we increase $K_1$ with step size $\Delta K_1$. In addition, {{in the swing-equation formulation, $\dot{\theta}$ represents the deviation of the oscillator frequency from the reference frequency in the rotating frame \cite{filatrella2008}}}; hence, it is drawn from a uniform distribution around zero $\dot{\theta} \in (-0.5, 0.5)$. For the backward direction, simulations are started from the synchronized state at a finite non-zero value of $K_1$, for which the initial condition is set as $\theta_i = 2\pi$ $\forall\,\, i \in \{1, 2, \dots, N\}$, whereas $\dot{\theta} \in (-0.5, 0.5)$ distributed uniformly. The final state from the previous $K_1$ is used as the initial condition for the next iteration.
%, except for the first value in both directions. 
We calculate $r_1$ and $r_2$ using Eq.~\eqref{order_par} by averaging over $8\times 10^4$ iterations after discarding an initial transient of $5\times 10^4$.
\paragraph{Adaptation induced transitions:}

Contrary to the analytical calculations determined using Eq.~\eqref{r1_eq} for ($\gamma=1$), numerical simulations of Eq.~\eqref{Mean_field} { for finite $N=500$} yield an abrupt jump in the forward direction depicted by the violet square in Fig.~\ref{1}(b). This abrupt jump occurs in a finite-size system due to fluctuations in determining $r_{1}$ (Eq.~\eqref{order_par}), which becomes meaningful in the bistable region where both incoherent and synchronized states are stable. The system will exhibit a jump to a synchronized state if the fluctuations cross the unstable state. { As an unstable solution, this state is not observed in direct numerical simulations.} However, the analytically determined lower branch solution of $r_1$ { for $\gamma=1$} (solid lines in Fig.~\ref{1}(b)) approaches the incoherent state asymptotically and provides insight into the mechanism of jumps to the synchronized state. When fluctuations in $r_1$ push the dynamics beyond this lower branch, a transition to a synchronization state occurs. { This underscores the interplay between adaptive coupling and the finite-size limit of oscillators in determining the nature of the synchronization transition.} Furthermore, in the forward direction starting with $K_1=0$, we increase it and note the value of coupling strength $K_{1f}$ at which, within a fixed time, the system exhibits the jump to the synchronized state from the incoherent state.
As expected, different realizations of uniform random initial conditions for phases { lead to slight variations in the magnitude of fluctuations of $r_1$ in the incoherent state. In the bistable regime, these fluctuations drive the system to cross the unstable branch over a range of $K_1$ values. Consequently, the forward transition point $K_{1f}$ is not strictly deterministic for finite $N$, but instead lies within a narrow interval, as shown in Fig.~\ref{1}(c-e). To quantify this variability, we analyze the probability distribution of $K_{1f}$ for different values of $\gamma$.} Here, $\gamma$ controls the strength of adaptive feedback in Eq.~\eqref{Mean_field}, and an increase in $\gamma$ results in weakening the coupling strength through which oscillators are interacting, yielding an overall increase in mean $K_{1f}$ as depicted in Fig.~\ref{1}(c-e).

\paragraph{Forward transition point and occurrence of multistability:} As shown in Fig.~\ref{2}(a), the forward transition point $K_{1f}$ shifts toward higher coupling values with increasing $N$. This shift results from the reduction of fluctuations in $r_1$, which scale as $\delta r_1 \sim \mathcal{O}\!\left(\frac{1}{\sqrt{N}}\right)$. Since the system exhibits a jump towards a weakly synchronized state when fluctuations in $r_1$ drive it across the unstable branch, the suppression of these fluctuations with increasing $N$ requires larger values of $K_1$ to induce synchronization, {as illustrated in Fig.~\ref{2}(a-b)}. Here, the term weakly synchronized state does not refer to small $r_1$ values \cite{tanaka1997}; rather, {it represents the upper branch solution of $r_1$  calculated from Eq.~\eqref{r1_eq} for $p \in (1,2)$ under the frequency limit corresponding to case~$\mathrm{(I)}$.} In this regime, and for $K_2 \neq 0$, solutions of integral $r_2^d$ in Eq.~\eqref{r1_eq} diverges, thus only upper branch solution of $r_1$ is obtained. The numerical simulations in the forward direction agree with the analytically obtained solution for the weakly synchronized state. Additionally, in the backward direction, a change in $N$ does not impact the transition point, Fig.~\ref{2}(a).
{ The adaptive coupling modifies the basin boundaries, such that incoherent and synchronized states coexist; hence, the finite-size fluctuations can induce an abrupt transition to synchronization. As $N \to \infty$ fluctuations in $r_1$ vanish, effectively suppressing the abrupt jump in the thermodynamic limit. In contrast, in the absence of adaptation, the forward critical coupling approaches a finite value as $N \to \infty$.}  

\begin{figure}[t!]
\vspace{0.5cm}
\begingroup
  \begin{tabular}{c}
\hspace{-0.5cm}
\includegraphics[width=0.35\textwidth]{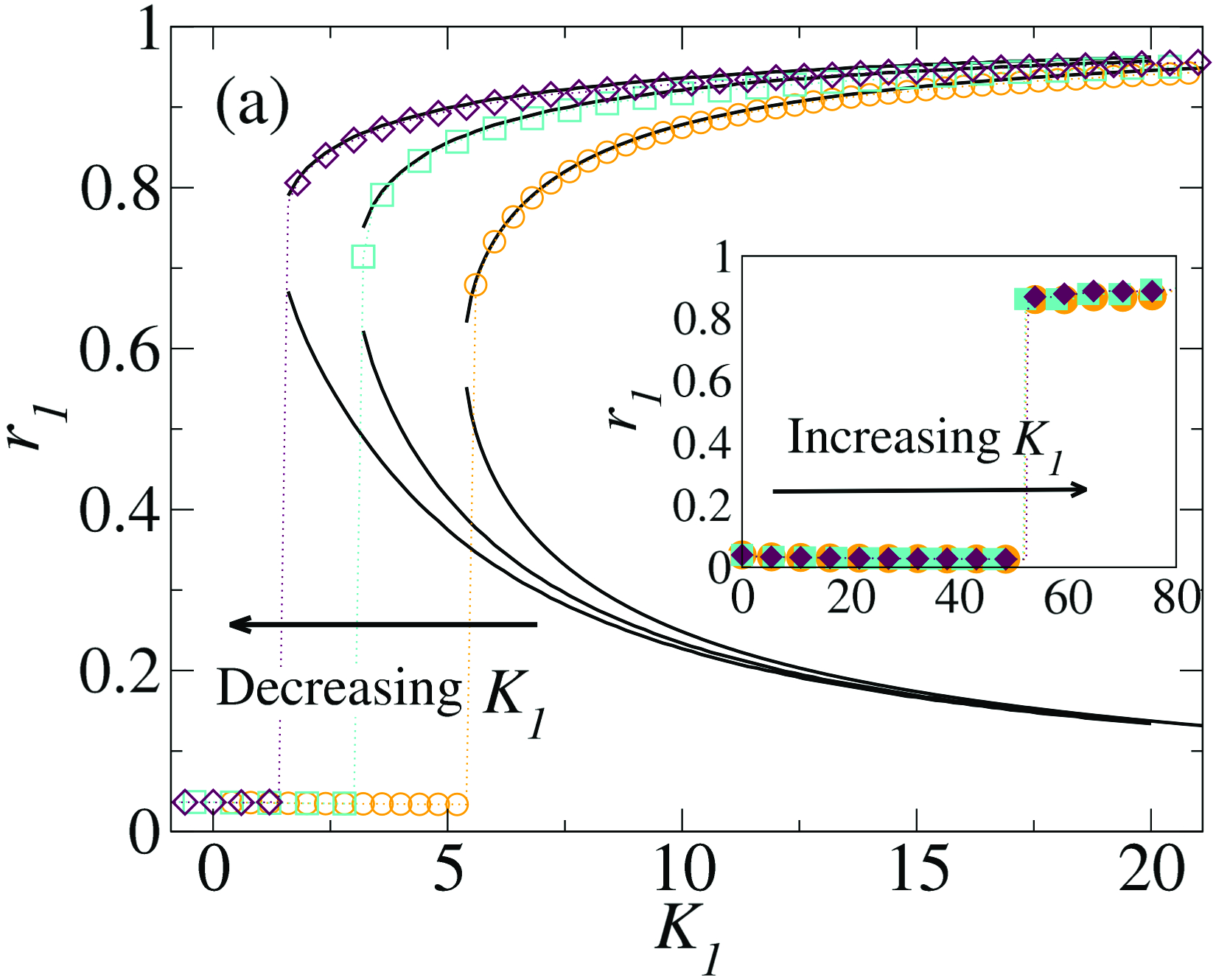}\\
%\hspace{0.05cm}
\includegraphics[width=0.352\textwidth]{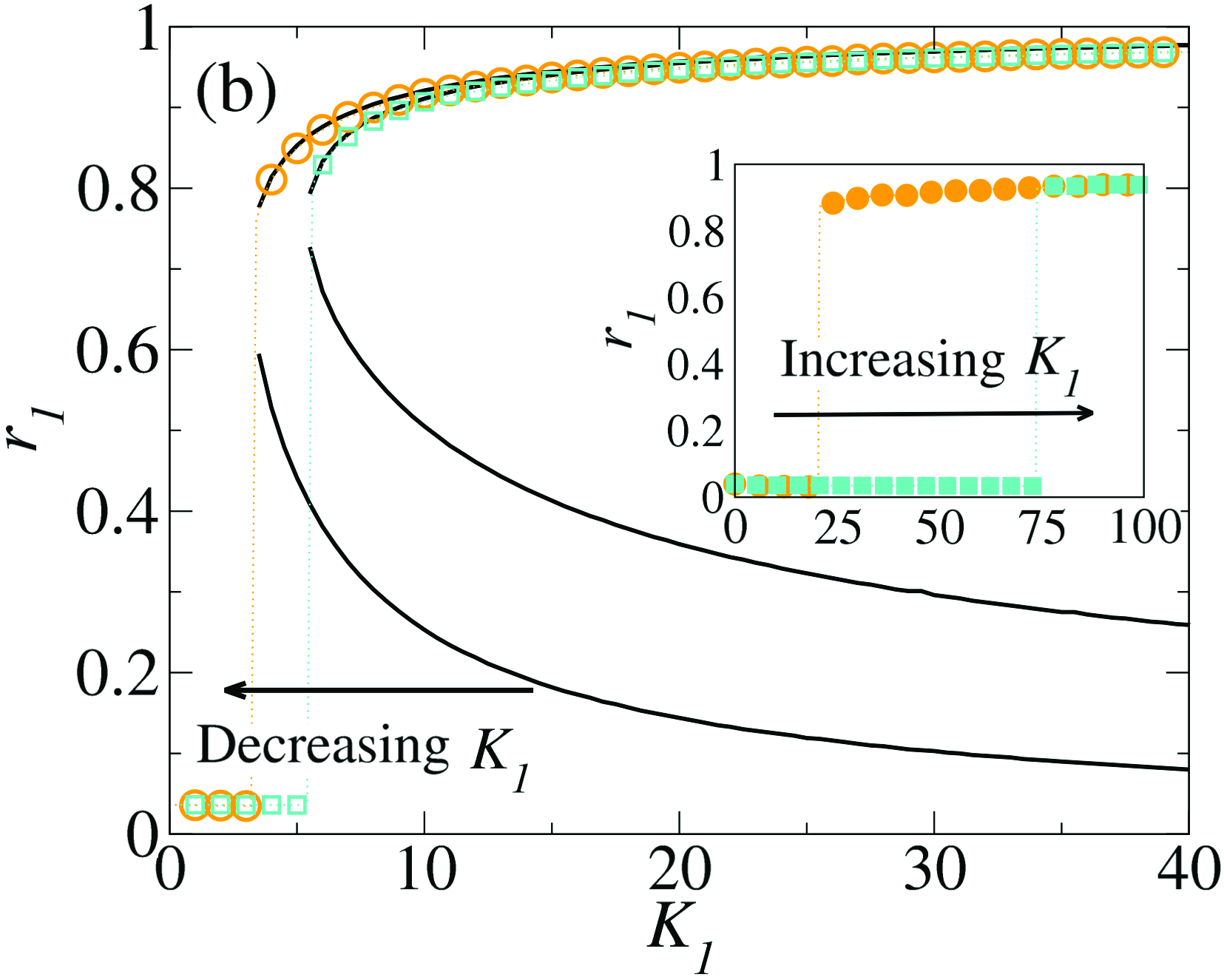} 
%\hspace{-0.28cm}
\end{tabular}
\endgroup
\caption{(Color online) Change in $K_{1b}$: $r_1$ vs $K_1$ for $N=500$ for backward direction numerical results obtained using (Eq.~\eqref{Mean_field}), and analytical results (solid lines) calculated for case $\mathrm{(II)}$ using Eqs.~\eqref{final_locked} and \eqref{final_drift}. (a) $m=3$ and $\gamma=1$ for $K_2=0$ (orange circle), $K_2=5$ (turquoise square), $K_2=8$ (maroon, diamond). (b) $m=1$ and $K_2=5$ for  $\gamma=1$ (orange circle), $\gamma=2$ (turquoise square). The insets display results from forward-direction numerical simulations.}
\label{3}
\end{figure}
Furthermore, an increase in mass requires a larger $K_1$ to exhibit a jump towards a weakly synchronized state (Fig.~\ref{2}(c)). Since the system dynamics is determined by $\ddot\theta=\frac{f(\theta,\dot\theta)}{m}$ where an increase in $m$ slows down the oscillations, in turn shifting $K_{1f}$ towards larger values \cite{tumash2018}. For a larger $m$, the range of intrinsic frequency of oscillators ($\Omega_\mathrm{I}$) participating in the locked state decreases. Consequently, $r_1$ attains lower values for larger $m$ as fewer oscillators will contribute to the locked state. Although for larger $m$, an increase in $K_1$ leads to an increase in $\Omega_\mathrm{I}$, eventually making the range of intrinsic frequency of oscillators similar; thus, we observe comparable values for $r_1$ for higher values of $K_1$ (Fig.~\ref{2}(c)). Further,  $r_1$ manifests a step-like structure in the synchronized state, evident at larger $m$ (Fig.~\ref{2}(c)) and finite $N=500$. Therefore, we simulate Eq.~\eqref{Mean_field} for $m=3$ in the forward direction up to a final value of $K_1$, and then decrease $K_1$ using the last forward configuration as the initial condition. This yields multiple branches of steady-state solutions for a finite-size system (Fig.~\ref{2}(d)). {All multistable states are confined between two bounding curves obtained from independent forward and backward simulations described above, consistent with the analytical predictions corresponding to the frequency limits of cases ($\mathrm{I}$) and ($\mathrm{II}$).}

\begin{figure}[t!]
%\vspace{-0.4cm}
\begingroup
 %\vspace{-0.5cm}
\begin{tabular}{cc}
\includegraphics[width=0.24\textwidth]{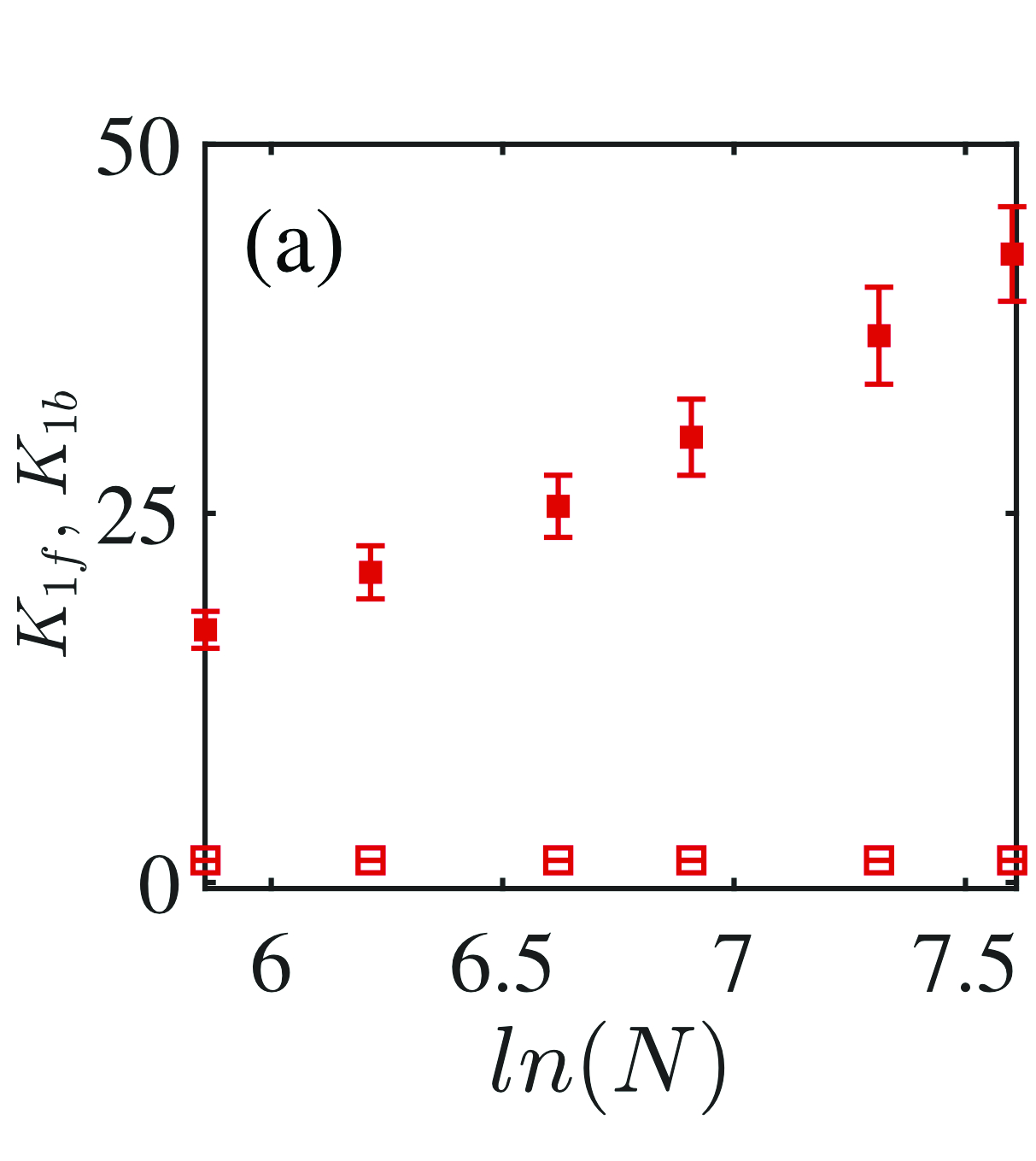}
\hspace{-0.25cm}
\vspace{-0.2cm}
\includegraphics[width=0.24\textwidth]{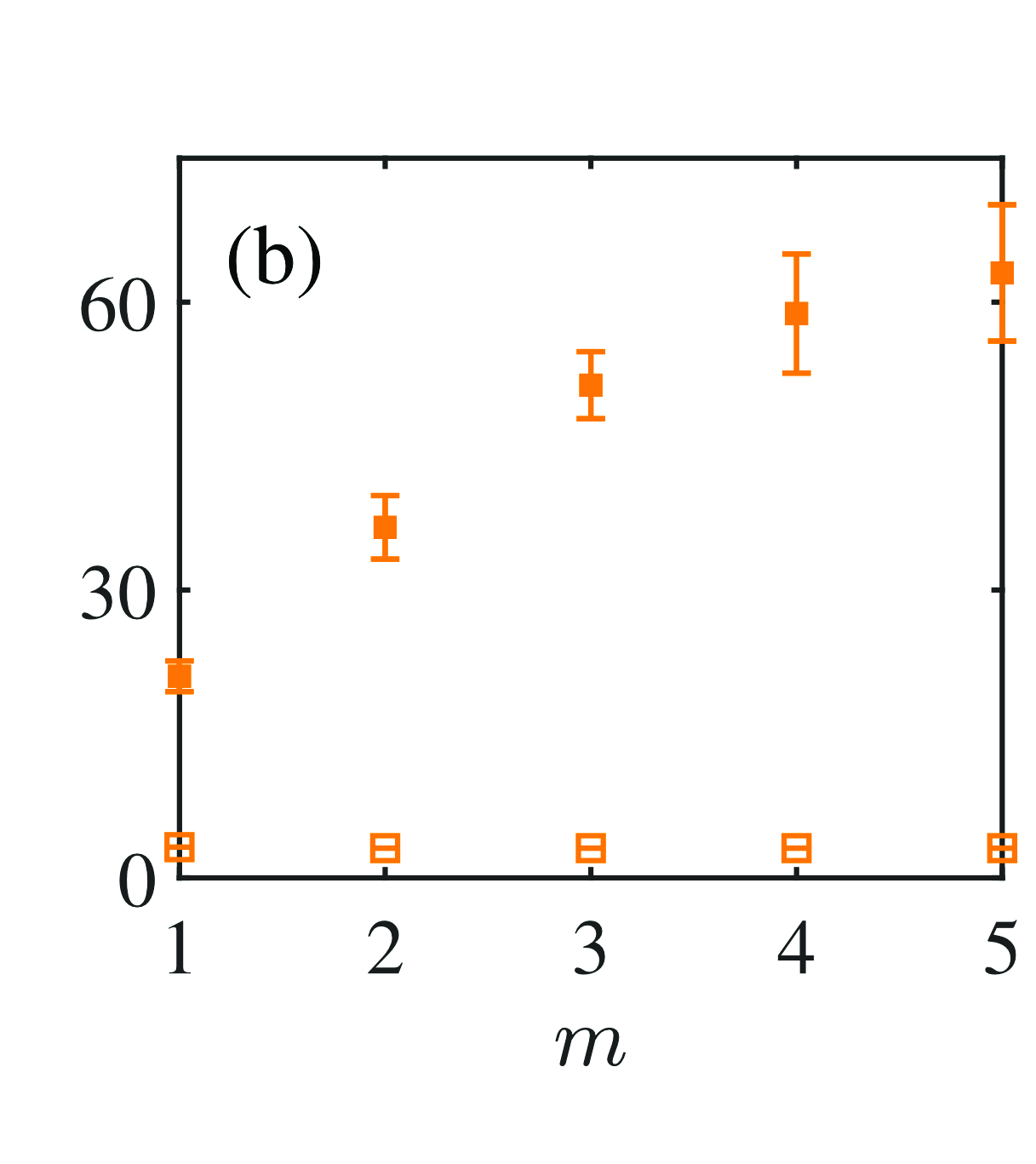}
\vspace{-0.2cm}
 %\vspace{} 
\end{tabular}
 %\vspace{-0.5cm} 
 \begin{tabular}{cc} 
\includegraphics[width=0.25\textwidth]{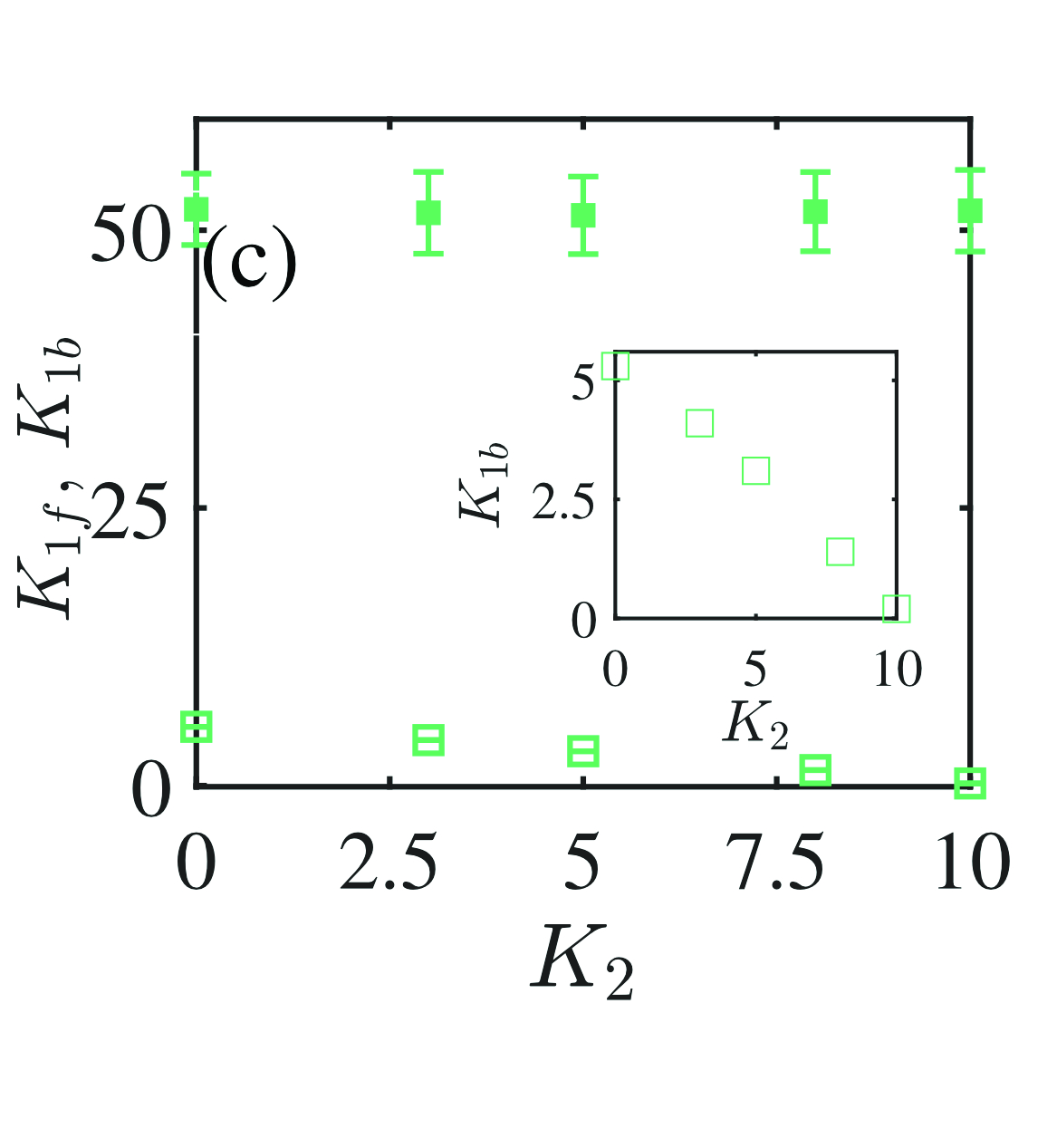} 
\hspace{-0.2cm}

\includegraphics[width=0.242\textwidth]{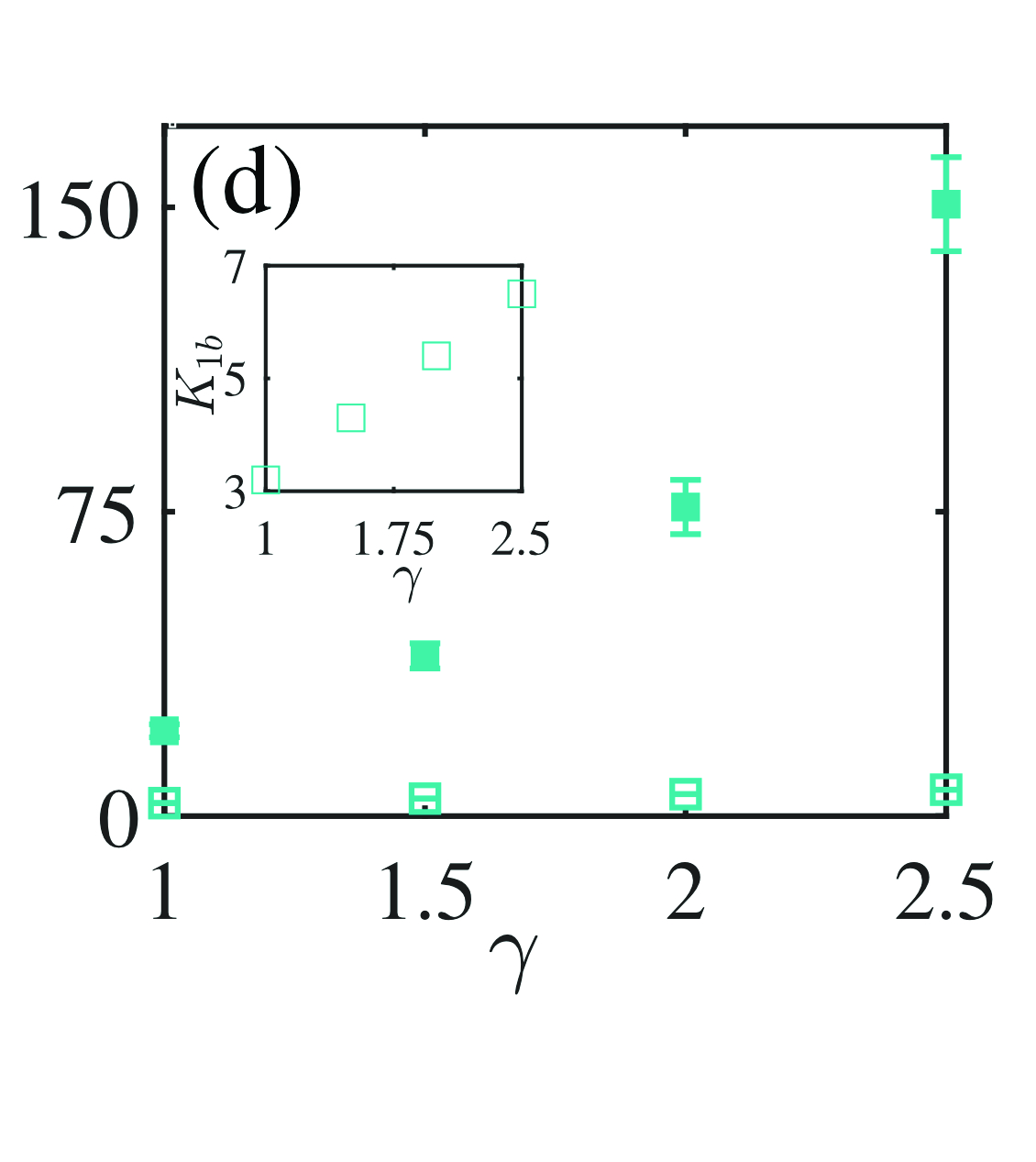} 
\end{tabular}
\endgroup
\vspace{-1cm}
\caption{Change in $K_{1f}$ and $K_{1b}$ as a function of (a) $N$ by fixing $m=1$, $\gamma=1$, and $K_2=8$,  (b) $m$ at fixed $N = 500$, $\gamma=1$, and $K_2 = 5$, (c)  $K_2$ at fixed  $N = 500$, $\gamma=1$, and $m = 3$, and (d) $\gamma$ for $N=500$, $m=1$, and $K_2=5$. Forward and backward points are represented by filled and open symbols, respectively.} %Solid lines indicate the linearly fitted curves.}
\label{4}
\end{figure}
\paragraph{Change in backward transition point:} A change in $K_2$ does not impact the forward transition point (inset), Fig.~\ref{3}(a). However, with an increase in $K_2$, $K_{1b}$ shifts to lower coupling values. Also, analytical calculation determined using Eq.~\eqref{r1_eq} for frequency limit $\Omega_\mathrm{II}<q$ represented by solid lines in  Fig.~\ref{3}(a) matches with the numerically obtained results. Notably, the adaptation exponent $\gamma$ dominates the effect of all other parameters and governs both the forward and backward transitions. %Fig.~\eqref{3}(b). 
As $\gamma$ increases, for fixed $K_2$, the backward transition point $K_{1b}$ shifts toward higher coupling values. {This behavior follows from the adaptive framework in Eq.~\eqref{Mean_field}, the factor $r_1^\gamma$ decreases with increasing $\gamma$, thereby reducing the coupling strength associated with pairwise and triadic interactions.} Consequently, the transition from the synchronized state to the incoherent state occurs at larger values of $K_{1b}$ (Fig.~\ref{3}(b)). Also, it shifts $K_{1f}$ towards a higher coupling value (inset, Fig.~\ref{3}(b))  obtained numerically for a finite-size system ($N=500$). For the backward direction, the analytical results using Eq.~\eqref{r1_eq} corresponding to the case ($\mathrm{II}$) are in full agreement with the numerically obtained results. 

\paragraph{Analysis of forward and backward transition points:} 
Since the value of $K_{1f}$ (at which the system exhibits an abrupt jump to a state in the weakly synchronized regime) changes with different realizations of initial conditions as described above, Fig.~\ref{1}(c-e). Therefore, we analyze the value of $K_{1f}$ considering $100$ realizations of uniform initial conditions of phases in the forward direction simulations of Eq.~\eqref{Mean_field}, and the mean and variance of $K_{1f}$ are evaluated to provide a probabilistic analysis (Fig.~\ref{4}). As $N$ and $m$ increase, the mean of $K_{1f}$ shifts towards larger coupling values; the bistable region also broadens, leading to an increased variance of $K_{1f}$, while $K_{1b}$ remains unaffected  (Fig.~\ref{4}(a-b)). Fig.~\ref{4}(c) illustrates that varying $K_2$ does not affect the mean of $K_{1f}$. However, an increase in $K_2$ shifts $K_{1b}$ toward lower values, thereby enlarging the bistable region and increasing the variance of $K_{1f}$ while keeping other system parameters fixed. Further, as $\gamma$ varies, there exists a change in both $K_{1f}$ and $K_{1b}$ towards higher values (Fig.~\ref{4}(d)). Also, as $\gamma$ increases, the variance in $K_{1f}$ increases. This analysis indicates that while system parameters such as $N$, $m$, and $K_2$ primarily influence either the forward or the backward transition point, the adaptive exponent $\gamma$ affects both. Moreover, $\gamma$ { controls the strength of adaptive feedback through $r_1^\gamma$ (Eq.~\eqref{Mean_field})}, thereby modulating the coupling strength and influencing how system parameters impact the transition points. Consequently, $\gamma$ plays a central role in determining the onset and stability of synchronization.

\begin{figure}[t!]
\vspace{0.5cm}
\includegraphics[width=0.35\textwidth]{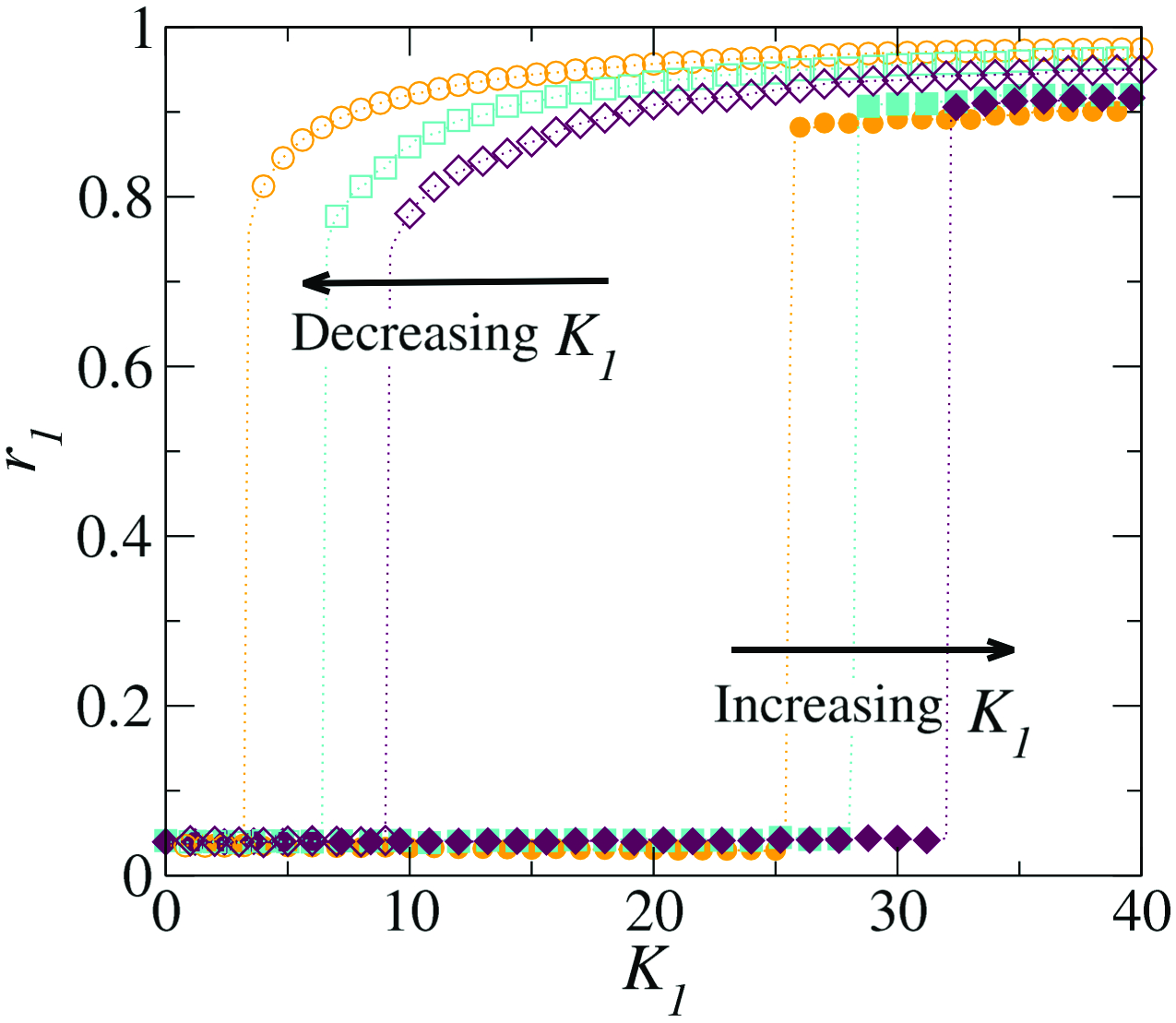} 
\caption{(Color online) Noise effect: $r_1$ vs $K_1$ for $m=1$, $K_2=5$ and $\gamma=1$ for different noise strength $D=0$ (orange circle), $D=0.5$ (violet square), $D=1$ (maroon diamond). Backward and forward directions are represented by open and filled symbols, respectively.}
\label{5}
\end{figure}

\paragraph{Effect of noise:} Introducing a white Gaussian noise $\xi(t)$ in Eq.~\eqref{GC_Model_Eq},
\begin{equation}\label{Noise_GC_Model_Eq}
\begin{split}
   m\ddot{\theta_i}=-\dot{\theta_i}&+{\Omega_i}+\frac{K_1r_{1}^{\gamma}}{N}\sum_{k=1}^{N}\sin(\theta_k-\theta_i)\\&+\frac{K_2r_{1}^{\gamma}}{N^2}\sum_{k=1}^{N}\sum_{l=1}^{N}\sin(2\theta_k-\theta_l-\theta_i)+ \xi_i(t),
 \end{split}   
\end{equation}
 with mean $\langle\xi_i(t)\rangle=0$ and covariance $\langle\xi_i(t)\xi_j(s)\rangle=2D\delta_{ij}\delta(t-s)$, where $D$ represents the noise strength. We simulate Eq.~\eqref{Noise_GC_Model_Eq} using Euler's method and analyze the impact of the role of the adaptive control mechanism on noise-induced transition points. When noise is incorporated with adaptive coupling, it requires a higher coupling strength to induce the abrupt jump in both forward and backward directions for fixed $m$ and $D$, Fig.~\ref{5}, as compared to the { without adaptive coupling} ($\gamma=0$) case in Eq.~\eqref{Noise_GC_Model_Eq} \cite{acebron2000, rajwani2025}. As $D$ increases, both forward and backward transition points shift towards larger coupling values for a fixed $\gamma$. Also, for $\gamma\neq 0$, an increase in $D$ causes the weakly synchronized state found in the forward direction simulations to gradually merge with the synchronized state obtained for the backward direction results (Fig.~\ref{5}).

\section*{Conclusion and Discussion}
We propose an analytically tractable extension of the second-order Kuramoto model that incorporates adaptive coupling strength with triadic ($2$-simplex) interactions {in a globally coupled network}. Theoretical calculations in the continuum limit predict that an incoherent state will persist as the pairwise coupling strength increases, {in contrast to previous studies without adaptive coupling, where a forward synchronization transition occurs even in the thermodynamic limit \cite{sabhahit2024}}. However, numerical simulations for a finite-size system indicate the forward transition within a bistable region. When finite-size fluctuations in the order parameter drive the system to surpass the unstable state, it results in a transition to a synchronized state. In fact, this leads to a range of transition points, which we determine by taking various realizations of uniform random initial conditions for the phases.  

Importantly, we show that  {the adaptation exponent $\gamma$ modulates the coupling strength of pairwise and triadic interactions through the factor $r_1^\gamma$}, which in turn governs both the forward and backward transition points. Additionally, we note that system size and inertia affect the forward transition point, whereas the $2$-simplex coupling strength shifts the backward transition point. {The adaptation exponent regulates how these parameters influence the transition points, highlighting the interplay between adaptive feedback, finite-size effect, inertia, and higher-order interactions in shaping collective dynamics.} Our findings are supported by analytical results obtained through the self-consistency method, and we highlight the presence of multistable states. Furthermore, incorporating perturbations such as white Gaussian noise enhances fluctuations in the system and shifts both the forward and backward transition points towards higher coupling values for a fixed adaptation exponent. 

While existing studies have primarily focused on fixed coupling strength, we extend the second-order Kuramoto model by implementing adaptive coupling, in which the coupling strength varies with the number of active oscillators participating in the synchronized state. This approach provides a {theoretical framework} for understanding feedback between collective dynamics and coupling strength in inertial oscillator systems. {Recent studies of adaptive second-order Kuramoto models on realistic power-grid networks have shown that the coupling strength can depend on the collective dynamical state to mitigate cascade failures \cite{benedek2025}, supporting the broader concept of feedback between synchronization dynamics and interaction strength.}
Moreover, the present work is {restricted to a globally coupled network and an analytically tractable adaptive form of coupling.} Extending this framework to real-world adaptive networks \cite{fialkowski2023} could enhance the understanding of transition points in such systems. This study can also be further extended by considering adaptation as a more generalized function of the order parameter \cite{jin2023}, which may provide a broader description of adaptive feedback mechanisms in complex dynamical networks.

\section*{Methods}\label{methods}
Here, we present details of the self-consistency method for the derivation of Eq.~\eqref{r1_eq}. In the continuum limit ($N \to \infty$),  the order parameter can be written as:
\begin{equation}
%\begin{split}
r_p = \int_{0}^{2\pi} \int_{-\infty}^{\infty} e^{\iota p \theta} \rho(\theta, \Omega) g(\Omega) \, d\Omega \, d\theta.
%r_2 &= \int_{0}^{2\pi} \int_{-\infty}^{\infty} e^{2i \theta} \rho(\theta, \Omega) g(\Omega) \, d\Omega \, d\theta.
%\end{split}
\end{equation}
In the steady state, coupled oscillators are described by a probability density function $\rho(\theta,\Omega)$, where  $\rho(\theta,\Omega)d\theta$ represents the fraction of oscillators with phases lying in the range $\theta$ to $\theta+d\theta$ for a given intrinsic frequency $\Omega$. The distribution function is then computed separately for oscillators locked to the mean phase and for those that are drifting; hence, the order parameter is $r_p=r_p^{l}+r_p^{d}$. Two different cases are considered: $\mathrm{(I)}$ corresponds to the case when one starts with the incoherent state at $K_1=0$ and $r_1=r_2=0$, and $\mathrm{(II)}$ indicates the case when one starts from a fully synchronized state. Furthermore, for $\gamma=0,\,\,K_2=0$ in Eq.~\eqref{GC_Model_Eq}, S. Olmi \textit{et. al.} describe how different frequency range affects the level of synchronization, and how $r_1$ varies with an increase or decrease of $K_1$  \cite{olmi2014}. In the first case, starting from an incoherent state, the oscillators are initially in a drifting state. Upon increasing $K_1$, the oscillators transition to a stable fixed point after a homoclinic bifurcation at $\beta \approx \frac{4}{\pi}\alpha$ calculated using Melnikov’s method \cite{tanaka1997, gao2018}. In the second case, starting from a fully synchronized state, the oscillators move to an incoherent state after the disappearance of the stable fixed point solution, which occurs for $\beta > 1$, Fig.~\eqref{1}(a). Hence, for the case $\mathrm{(I)}$, the frequency range is given by  $|\Omega_\mathrm{I}| < \frac{4}{\pi} \sqrt{\frac{q}{m}}$, and for the case $\mathrm{(II)}$, it is $|\Omega_\mathrm{II}|< q$.  

Furthermore, the contribution of the locked oscillators, by setting $\ddot{\theta} = \dot{\theta} = 0$, gives $\Omega = q\sin(\theta^*)$ (Eq.~\eqref{Mean_field}). In continuum limit the order parameter is defined as: $r_p^{l}=\int_{0}^{2\pi}  \int_{-{\Omega_{\mathrm{I}}/\Omega_{\mathrm{II}}}}^{{\Omega_{\mathrm{I}}/\Omega_{\mathrm{II}}}} e^{i p\theta} \delta(\theta-\theta^*) g(\Omega) \, d\Omega, d\theta.$ Since $g(-\Omega) = g(\Omega)$, the imaginary part becomes zero. Hence, the contribution is:
\begin{equation}\label{final_locked}
\begin{split}
    r_1^{l}&=  \int_{|\Omega|<\Omega_{\mathrm{I}}/\Omega_\mathrm{II}} \sqrt{1-\left({\frac{\Omega}{q}}\right)^2}g(\Omega) \, d\Omega,\\
    r_2^{l}&= \int_{|\Omega|<\Omega_{\mathrm{I}}/\Omega_\mathrm{II}} \left({1-2\left({\frac{\Omega}{q}}\right)^2}\right)g(\Omega) \, d\Omega.\\
\end{split}    
\end{equation}
Next, by using the method described by Gao and Efstathiou \cite{gao2018} we calculate contribution from the drifting oscillators as $r_{p}^d= \int_{|\Omega|>\Omega_{\mathrm{I}}/\Omega_\mathrm{II}}\int_{-\pi}^{\pi} e^{\iota p\theta}\rho_d(\theta,\Omega)g(\Omega) \, d\Omega$. Considering that drifting oscillators form a stationary distribution on a circle, we have  $\rho_d(\theta,\Omega) \propto \frac{1}{|\dot\theta|}$ which follows $\int_{-\pi}^{\pi}\rho_d(\theta,\Omega) d\theta\,=\, \int_{0}^{T}\rho_d(\theta,\Omega) \dot\theta \, dT=1$, where $T$ is the time period of limit cycle. Further, the contribution from drifting oscillators is given as:
\begin{equation}\label{r_drift}
    r_p^{d}=\int_{|\Omega|>\Omega_{\mathrm{I}}/\Omega_\mathrm{II}}\frac{1}{T}\int_{0}^{2\pi}\frac{e^{\iota p\theta}}{\dot\theta} d\theta \,\,g(\Omega) d\Omega.
\end{equation}
{It is worth noting that when mixed harmonics are considered in Eq.~\ref{GC_Model_Eq}, such as the pairwise interaction $\sin(\theta_j-\theta_i)$ together with the triadic term $\sin(\theta_k+\theta_j-2\theta_i)$, the resulting mean-field equation in Eq.~\eqref{Mean_field} cannot be reduced exactly to the form of Eq.~\eqref{time_MF}. As a consequence, the calculation of the frequency limit and the limit-cycle solution required to evaluate $r_p^{d}$ becomes challenging.
}
In the present work, we calculate $r_p^{d}$ by following Ref.~\cite{gao2018}, where an approximate limit-cycle solution of Eq.~\eqref{time_MF} is obtained by representing $\dot\theta$ as a function of $\theta$ for points in the limit cycle using a Fourier series: $\dot\theta=A_0+A_1 \cos\theta+B_1\sin\theta$.  Further, substituting in Eq.~\eqref{time_MF} and comparing the coefficients, while ignoring the higher harmonic terms, yields the following:
\begin{equation}\label{Eq_LC}
    \dot\theta=\nu_0+\delta \cos(\theta+\theta^*),
\end{equation}
where $\nu_0=\frac{\beta}{\alpha}$ and $\frac{1}{\delta}e^{\iota\theta^*}=\nu_0+\iota\alpha$. Considering $\theta$ as $\nu_0 t$ and integrating Eq.~\eqref{Eq_LC} with respect to time gives $\theta(t,-\Omega)=-\theta(t,\Omega)$; therefore,  contribution from the imaginary part in Eq.~\eqref{r_drift} goes to zero. Consequently, the contribution of drifting oscillators is
\begin{equation}\label{final_drift}
    r_p^{d}=\int_{|\Omega|>\Omega_{\mathrm{I}}/\Omega_\mathrm{II}} \langle \cos (p\theta) \rangle g(\Omega) d\Omega,
\end{equation}
where $\langle e^{\iota p\theta} \rangle=\frac{1}{T} \int_{0}^{2\pi}\frac{e^{\iota p\theta}}{\dot\theta}d\theta$. For $p \in \{1, 2\}$, expressions for $\langle \cos(p\theta)\rangle$ are 
\begin{equation}\nonumber
    \begin{split}
    \langle \cos\theta\rangle&=-\nu_0^2+\nu_0\sqrt{\nu_0^2-\delta^2}\\
    \langle \cos2\theta\rangle&=(\nu_o^2\delta^2-\alpha^2\delta^2)\left(\frac{2\nu_0}{\delta^2}\left(\nu_0-\sqrt{\nu_0^2-\delta^2}\right)-1\right).
    \end{split}
\end{equation}
Finally, we calculate the expression for $r_p=r_p^l+r_p^d$ for $p\in\{1,2\}$.
%using Eqs.~\eqref{final_locked} and ~\eqref{final_drift}.

\section*{Data availability}
All the data supporting the findings of this study are available within the paper.
\section*{Code availability}
All codes used for this study are available from the corresponding author upon request. 
%\newpage

%\begin{figure}[t!]
 %   \centering
%\includegraphics[width=0.45\textwidth]{Bi_N_500_al1_m1_K2.eps}
 %   \caption{(Color online) $r_1$ vs $K_1$ }
  %  \label{}
%\end{figure}
\section*{Acknowledgment}
SJ and PR acknowledge the Govt of India SERB Power grant SPF/2021/000136 and PMRF grant PMRF/2023/2103358, respectively. We thank Mehrnaz Anvari for discussions on relevance of the model for power grid systems, and Benjamin Sch\"afer for insightful comments and suggestions. 
%\section*{Author contributions}
\section*{Competing interests}
The authors declare no competing interests.
%\section*{Additional information}

\begin{comment}
\begin{figure}[t!]
\begingroup
 % \begin{tabular}{cc}
\includegraphics[width=0.45\textwidth]{Noise_d_N_2000_al_1_m1_k2_5.eps} 

%\includegraphics[width=0.24\textwidth]{N_500_al_1_m_k2_5.eps} 
%\end{tabular}
\endgroup
\caption{(Color online) } 
\label{}
\end{figure}

\begin{figure}[t!]
\begingroup
 % \begin{tabular}{cc}
\includegraphics[width=0.45\textwidth]{Noise.eps} 

%\includegraphics[width=0.24\textwidth]{N_500_al_1_m_k2_5.eps} 
%\end{tabular}
\endgroup
\caption{(Color online) } 
\label{}
\end{figure}
\end{comment}
\end{document}